# Ab Initio Second-Order Nonlinear Optics in Solids: Second-Harmonic Generation Spectroscopy from Time-Dependent Density-Functional Theory

Eleonora Luppi, Hannes Hübener, and Valérie Véniard

*Laboratoire des Solides Irradiés, École Polytechnique,*

*CNRS-CEA/DSM, F-91128 Palaiseau,*

*European Theoretical Spectroscopy Facility (ETSF), France*

(Dated: June 11, 2010)

## Abstract

We present in detail the formulation of the *ab initio* theory we have developed for the calculation of the macroscopic second-order susceptibility $\chi^{(2)}$. We find a general expression for $\chi^{(2)}$ valid for any fields, containing the *ab initio* relation between the *microscopic* and *macroscopic* formulation of the second-order responses. We consider the long wavelength limit and we develop our theory in the Time-Dependent Density-Functional Theory framework. This allows us to include straightforwardly many-body effects such as crystal local-field and excitonic effects. We compute the Second-Harmonic Generation spectra for the cubic semiconductors SiC, AlAs and GaAs and starting from the Independent-Particle Approximation for $\chi^{(2)}$, we include quasiparticle effects via the scissors operator, crystal local-field and excitonic effects. In particular, we consider two different types of kernels: the ALDA and the "long-range" kernel. We find good agreement with other theoretical calculations and experiments presented in literature, showing the importance of very accurate description of the many-body interactions.

## I. INTRODUCTION

Nonlinear optics is an important and exciting field of fundamental and applied research, with applications in many different disciplines like materials science[1], chemistry[2,3] and biology[4]. Among all the nonlinear phenomena existing in nature, a major role is played by the second-order process: Second-Harmonic Generation (SHG).

Since its discovery in 1961[5] the importance of SHG has grown, because of its sensitivity to space symmetry, making SHG an extremely versatile tools for studying many kinds of surfaces[6–10], superlattices[11] and interfaces[10,12–14]. Nowadays this technique is also used for characterizing systems like interfaces of nanocrystals[15] or as a probe for molecular chirality in polymers[16,17] and nanotubes[18]. Furthermore, SHG is also interesting for the development of optoelectronic devices. Many experimental efforts are made towards the design, fabrication, and search for new nonlinear optical materials and SHG techniques play a central role in these studies[19,20].

In the SHG nonlinear optical[21,22] process the interaction of matter with light is described by the macroscopic second-order susceptibility $\chi^{(2)}$. This quantity includes the many-body interactions between the electrons of the system: the variation of the screening fields on the microscopic scale, i.e. crystal local-field effects[23] and the electron-hole interaction, i.e. excitonic effects[24], when real and/or virtual excitations are created in the process. The basic requirement of such a description is a comprehensive understanding of the nonlinear microscopic physical mechanisms in the second-order response and the corresponding macroscopic relation with physical measurable quantities. This is a formidable task and considerable difficulties have delayed accurate calculations for many years.

In the first theoretical works on nonlinear optics[25–27], the second-order nonlinear optical susceptibility was calculated for finite- and zero-frequency values in zinc-blende crystals and despite the fitting parameters used[26,27] a comparison between theoretical and experimental data was difficult. The empirical pseudopotential calculation of Fong and Shen[26] underestimated the experimental values by 1-2 orders of magnitude, while Moss et *al.*[27] overestimated the values of the $\chi^{(2)}$ by a factor 1-4.

Later, Levine and coworkers[28–33] presented an ab initio formalism for the calculation of the second-harmonic susceptibility in solids, performed in the context of the one-electron band theory which takes into account crystal local-field effects. Sipe and Ghahramani[34] and Aversa and Sipe[35] developed a formalism for the calculation of the second-order optical response of crystals in the independent particle approximation, and a more recent approach has been reported by Sipe and Shkrebtii[36]. Based on these works, Hughes and Sipe[37–39] presented a first-principles calculation of the second-order optical response functions, including self-energy corrections at the level of the scissors approximation. A different approach from the sum over states methods cited above, was presented by Dal Corso *et al.*[40], in which self-consistent local-field effects were included in $\chi^{(2)}$ within local density approximation (LDA) through the "2n+1" theorem in the Time-Dependent Density-Functional Theory (TDDFT) framework, and applied to semiconductor materials[41] and molecules[42].

Most of these works focus on the calculation of the optical second-harmonic susceptibility, in the static limit and in the low-frequency range. Theoretical analysis were also extended to a larger frequency range in particular in the Independent Particle Approximation (IPA)[37–39,43,44]. These calculations improve the description of the second-order susceptibility, however, the details of each approach show clearly that the calculation of the second-order optical susceptibility still remains a nontrivial task, and the same accuracy obtained nowadays in linear optics has not been achieved yet, in particular for a large frequency range.

Furthermore, only a few works exist on the inclusion of excitonic effects in $\chi^{(2)}$. Chang et al.[45] and Leitsman et al.[46] presented an ab initio many-body formalism for computing the frequency-dependent second-harmonic polarizability of semiconductor materials including crystal local-field and excitonic effects. In their method the electron-hole interaction is described through the solution of an effective two-particle Hamiltonian, derived from the Bethe-Salpeter equation (BSE), which has been successfully used for linear optics. The authors found a reasonable agreement with experimental data in the static limit, while in a larger energy range the comparison was not satisfactory. The question arises whether it is possible to use this method to describe excitons for higher-order calculations.

In this work, we present in detail the formalism we developed and presented in Ref.[47] for the calculation of the macroscopic second-order susceptibility $\chi^2$, valid for crystals and molecular systems. We find an expression for $\chi^2$ valid for any kind of classical field (longitudinal and transverse) which relates the microscopic and macroscopic formulations of second-order response. We have applied our theory to SHG spectroscopy, considering only the case of vanishing light wave vector ($q \rightarrow 0$). In this limit, the macroscopic second-order susceptibility $\chi^2$ can be expressed in the TDDFT[48–50] framework. Within our approach crystal local-field and excitonic effects are straightforwardly included in the macroscopic second-order susceptibility.

The paper is organized as follows: in Sec. II, we present the derivation of the relation between the microscopic and macroscopic formulation of the second-order response, obtaining a general expression for the macroscopic second-order susceptibility $\chi^2$ valid for any fields. In Sec. III, we rewrite the general expression for $\chi^2$ as obtained from the previous section, for longitudinal fields (long wavelength limit), which permits us to write our formalism in the TDDFT framework. Furthermore, since in our approach we explicitly consider the symmetry properties of the system in order to obtain a specific component of the tensor $\chi^{(2)}$, we give the expression for the component $\chi^{(2)}_{xyz}$ for cubic symmetry. In Sec. IV we apply our method to the calculation of SHG spectroscopy for the cubic semiconductors SiC, AlAs and GaAs. We discuss SHG spectra obtained within different levels of description of the many-body interactions. Starting from the Independent-Particle Approximation for $\chi^{(2)}$, we include quasiparticle effects via the scissors operator, crystal local-field and excitonic effects.

## II. MACROSCOPIC SECOND-ORDER SUSCEPTIBILITY

In this section, we present in detail the derivation[47] of the macroscopic second-order susceptibility tensor $\chi^{(2)}$, which is defined by

$$\mathbf{P}^{(2)}_M = \chi^{(2)} \mathbf{E}_M \mathbf{E}_M, \tag{1}$$

$\mathbf{P}^{(2)}_M$ is the macroscopic second-order polarization and $\mathbf{E}_M$ the macroscopic component of the total electric-field. $\chi^{(2)}$ describes the response of a system to an external perturbation, containing the physical informations of the nonlinear microscopic response of the material, together with its many-body interactions. Moreover, as $\chi^{(2)}$ is a macroscopic quantity, it is directly related to measurable quantities, like SHG spectra.

In our formalism, in order to obtain $\chi^{(2)}$, we calculate in a first step the second-order *microscopic* polarization of the system, from which we derive in a second step the second-order *macroscopic* polarization ($\mathbf{P}^{(2)}_M$) of the system. Once we obtain an expression for $\mathbf{P}^{(2)}_M$, we derive an expression for the second-order susceptibility $\chi^{(2)}$. In this way, we obtain a general relation between *microscopic* and *macroscopic* formulation of the second-order response function, valid for any fields (longitudinal and/or transverse) and for any symmetry of the system.

### A. First Step: Microscopic Second-Order Polarization

We consider the response of a system of electrons, interacting via the Coulomb potential, perturbed by an external electromagnetic field. To calculate the second-order microscopic polarization of this system, we extend to second-order the linear response formalism of Del Sole and Fiorino presented in Ref.[51].

Using second-order time-dependent perturbation theory, we calculate the induced current $\mathbf{J}_{ind}$ as a function of the perturbing field $\mathbf{E}^p$, defined as[51]

$$\mathbf{E}^p = \mathbf{E}^{ext} + \mathbf{E}^{i,T} = \mathbf{E} - \mathbf{E}^{i,L}. \tag{2}$$

where $\mathbf{E}^{ext}$ is the external applied electric-field, $\mathbf{E}$ is the total microscopic field and $\mathbf{E}^{i,L}$ and $\mathbf{E}^{i,T}$ are respectively the longitudinal and transverse components of the field induced by the external perturbation.

The relation between the polarization and the induced current is

$$\mathbf{P}(\mathbf{r},t) = \int_{-\infty}^{t} \mathbf{J}_{ind}(\mathbf{r},t')dt', \tag{3}$$

and from its time Fourier transform we obtain an expression for the first[51,52]

$$\mathbf{P}^{(1)}(\mathbf{r},\omega) = \int d\mathbf{r}_1 \tilde{\alpha}^{(1)}(\mathbf{r},\mathbf{r}_1,\omega)\mathbf{E}^P(\mathbf{r}_1,\omega), \tag{4}$$

and for the second-order microscopic polarization

$$\mathbf{P}^{(2)}(\mathbf{r},\omega) = \int d\mathbf{r}_1 d\mathbf{r}_2 \int d\omega_1 d\omega_2 \delta(\omega - \omega_1 - \omega_2)$$
$$\tilde{\alpha}^{(2)}(\mathbf{r},\mathbf{r}_1\mathbf{r}_2,\omega_1,\omega_2)\mathbf{E}^P(\mathbf{r}_1,\omega_1)\mathbf{E}^P(\mathbf{r}_2,\omega_2). \tag{5}$$

The quantity $\tilde{\alpha}^{(1)}$ (Eq. (4)) is the linear quasi-polarizability tensor[52] defined in terms of the current-current response function $\chi_{\mathbf{jj}}$ (see Appendix A) as

$$\tilde{\alpha}^{(1)}(\mathbf{r},\mathbf{r}_1,\omega) = \frac{1}{\omega^2}\{\chi_{\mathbf{jj}}(\mathbf{r},\mathbf{r}_1,\omega) - \langle\hat{\rho}(\mathbf{r})\rangle\delta(\mathbf{r}-\mathbf{r}_1)\} \tag{6}$$

and the quantity $\tilde{\alpha}^{(2)}$ (Eq. (5)) is the second-order quasi-polarizability defined in terms of the current-current-current response function $\chi_{jjj}$ (see Appendix A) as

$$\tilde{\alpha}^{(2)}(\mathbf{r}, \mathbf{r}_1, \mathbf{r}_2, \omega_1, \omega_2) = \frac{-i}{\omega_1 \omega_2 (\omega_1 + \omega_2)}$$
$$\left\{ \frac{1}{2} \chi_{jjj}(\mathbf{r}, \mathbf{r}_1, \mathbf{r}_2, \omega_1, \omega_2) - \chi_{\rho j}(\mathbf{r}, \mathbf{r}_1, \omega_1) \delta(\mathbf{r} - \mathbf{r}_2) \right.$$
$$\left. - \chi_{j\rho}(\mathbf{r}, \mathbf{r}_1, \omega_1 + \omega_2) \delta(\mathbf{r}_1 - \mathbf{r}_2) \right\}, \tag{7}$$

written in atomic units ($\hbar = 1$, $e = 1$ and $m = 1$). We note that the quantities $\tilde{\alpha}^{(1)}$ and $\tilde{\alpha}^{(2)}$ are called quasi-polarizability because they are related to $\mathbf{E}^p$, while the true polarizabilities are related to $\mathbf{E}$.

Performing the space Fourier transform of Eq. (4) and Eq. (5), we obtain

$$\mathbf{P}_\mathbf{G}^{(1)}(\mathbf{q}, \omega) = \sum_{\mathbf{G}_1} \tilde{\alpha}^{(1)}_{\mathbf{G}, \mathbf{G}_1}(\mathbf{q}, \mathbf{q}, \omega) \mathbf{E}^P_{\mathbf{G}_1}(\mathbf{q}, \omega), \tag{8}$$

$$\mathbf{P}_\mathbf{G}^{(2)}(\mathbf{q}, \omega) = \sum_{\substack{\mathbf{q}_1, \mathbf{q}_2, \\ \mathbf{G}_1, \mathbf{G}_2}} \int d\omega_1 d\omega_2 \delta_{\mathbf{q}, (\mathbf{q}_1 + \mathbf{q}_2)} \delta(\omega - \omega_1 - \omega_2) \times$$
$$\tilde{\alpha}^{(2)}_{\mathbf{G}, \mathbf{G}_1, \mathbf{G}_2}(\mathbf{q}, \mathbf{q}_1, \mathbf{q}_2, \omega_1, \omega_2) \mathbf{E}^p_{\mathbf{G}_1}(\mathbf{q}_1, \omega_1) \mathbf{E}^p_{\mathbf{G}_2}(\mathbf{q}_2, \omega_2), \tag{9}$$

where $\mathbf{q}$ is a vector in the first Brillouin zone, $\mathbf{G}$, $\mathbf{G}_1$ and $\mathbf{G}_2$ are vectors of the reciprocal lattice and $\mathbf{P}_\mathbf{G}(\mathbf{q}, \omega)$ stands for $\mathbf{P}(\mathbf{q} + \mathbf{G}, \omega)$.

We obtain in Eq. (8) and in Eq. (9) the first[51,52] and second-order microscopic polarizability as a function of the perturbing electric field $\mathbf{E}^p$ defined in Eq. (2). We will use both these polarizations in the following, in order to derive an expression for the macroscopic $\chi^{(2)}$.

### B. Second Step: Macroscopic Second-Order Polarization

The physical properties of a system are described by macroscopic quantities. For instance, the linear optical properties of a material are obtained through the macroscopic dielectric tensor $\epsilon_M$ defined through the relation $\mathbf{D}_M = \epsilon_M \mathbf{E}_M$ where $\mathbf{D}_M$ is the macroscopic electric displacement vector in the linear response and $\mathbf{E}_M$ is the macroscopic component of the total electric field[93]. In the second-order response, the quantity which describes the optical properties of the system is the macroscopic second-order susceptibility $\chi^{(2)}$, defined in Eq.(1).

$\epsilon_M$ and $\chi^{(2)}$ can not be directly obtained from the microscopic quantities defined in Eq. (8) and Eq. (9), as both $\epsilon_M$ and $\chi^{(2)}$ are written in terms of the macroscopic polarization which depends on the total electric field, containing the external perturbation and the induced response of the system to the perturbation. The microscopic quantities in Eq. (8) and Eq. (9) have to be expressed as a function of the total electric field and they have to be spatially averaged[52]. In practice, for a quantity written in reciprocal space, the averaging procedure consists in keeping only the $\mathbf{G} = \mathbf{0}$ component[52].

In order to find an expression for the macroscopic second-order susceptibility $\chi^{(2)}$ (Eq.(1)) we start to consider the microscopic electric displacement vector where we expand the polarization up to second-order

$$\mathbf{D}_\mathbf{G}(\mathbf{q},\omega) = \mathbf{E}_\mathbf{G}(\mathbf{q},\omega) + 4\pi(\mathbf{P}_\mathbf{G}^{(1)}(\mathbf{q},\omega) + \mathbf{P}_\mathbf{G}^{(2)}(\mathbf{q},\omega)). \tag{10}$$

To obtain the macroscopic electric displacement, we need to spatially average[52] $\mathbf{P}_\mathbf{G}^{(1)}$ and $\mathbf{P}_\mathbf{G}^{(2)}$, given in Eq. (8) and Eq. (9), and to express these polarizations in terms of the total electric field. In Eq. (8) and Eq. (9) the microscopic polarizations are expressed as functions of $\mathbf{E}_\mathbf{G}^p$, which significantly facilitates the calculation of the macroscopic averages. In fact, if the applied external field is of long wavelength, the perturbing field $\mathbf{E}_\mathbf{G}^p(\mathbf{q},\omega)$ becomes macroscopic $\mathbf{E}_\mathbf{G}^p(\mathbf{q},\omega) = \delta_{\mathbf{G},\mathbf{0}} \mathbf{E}_\mathbf{0}^p(\mathbf{q},\omega)$, as shown in Ref.[51] and we obtain from Eq. (8)

$$\mathbf{P}_\mathbf{0}^{(1)}(\mathbf{q},\omega) = \tilde{\alpha}_{0,0}^{(1)}(\mathbf{q},\mathbf{q},\omega) \mathbf{E}_\mathbf{0}^P(\mathbf{q},\omega). \tag{11}$$

To express Eq. (11) as a function of the total electric field, we use the relation between the perturbing and the total field obtained from Maxwell's equations in Ref.[51], which reads as

$$\mathbf{E}_\mathbf{G}^p(\mathbf{q},\omega) = \mathbf{E}_\mathbf{G}(\mathbf{q},\omega) + 4\pi \frac{\mathbf{q}+\mathbf{G}}{|\mathbf{q}+\mathbf{G}|} P_\mathbf{G}^L(\mathbf{q},\omega), \tag{12}$$

where $P_\mathbf{G}^L(\mathbf{q},\omega) = \frac{\mathbf{q}+\mathbf{G}}{|\mathbf{q}+\mathbf{G}|} P_\mathbf{G}(\mathbf{q},\omega)$ is the microscopic longitudinal polarization.

Inserting the spatial macroscopic average of Eq. (12) in Eq. (11), we obtain

$$\mathbf{P}_\mathbf{0}^{(1)}(\mathbf{q},\omega) = \tilde{\alpha}_{0,0}^{(1)}(\mathbf{q},\mathbf{q},\omega) \mathbf{E}_\mathbf{0}(\mathbf{q},\omega) + 4\pi \tilde{\alpha}_{0,0}^{(1)}(\mathbf{q},\mathbf{q},\omega) \times$$
$$\frac{\mathbf{q}}{q} \frac{\mathbf{q}}{q} (\mathbf{P}_\mathbf{0}^{(1)}(\mathbf{q},\omega) + \mathbf{P}_\mathbf{0}^{(2)}(\mathbf{q},\omega)) \tag{13}$$

where $q = |\mathbf{q}|$.

Note that Eq. (11) is a linear relation between two first order quantities, while the relation between the perturbed and the total fields in Eq. (12), contains implicitly higher order terms. The crucial point here is that, although the relation in Eq. (11) is linear in terms of the pertubing field, it will contain higher order terms when expressed in terms of the total field.

We solve Eq. (13) for $\mathbf{P}_0^{(1)}(\mathbf{q},\omega) + \mathbf{P}_0^{(2)}(\mathbf{q},\omega)$ obtaining

$$\mathbf{P}_0^{(1)}(\mathbf{q},\omega) + \mathbf{P}_0^{(2)}(\mathbf{q},\omega) = A(\mathbf{q},\omega)\tilde{\alpha}_{0,0}^{(1)}(\mathbf{q},\mathbf{q},\omega)\mathbf{E}_0(\mathbf{q},\omega) + A(\mathbf{q},\omega)\mathbf{P}_0^{(2)}(\mathbf{q},\omega) \qquad (14)$$

where

$$A(\mathbf{q},\omega) = \left[1 - 4\pi\frac{\mathbf{q}\,\mathbf{q}}{q\,q}\tilde{\alpha}_{0,0}^{(1)}(\mathbf{q},\mathbf{q},\omega)\right]^{-1} = 1 + 4\pi\frac{\mathbf{q}\,\mathbf{q}}{q\,q}\frac{\tilde{\alpha}_{0,0}^{(1)}(\mathbf{q},\mathbf{q},\omega)}{1 - 4\pi\tilde{\alpha}_{0,0}^{(1),LL}(\mathbf{q},\mathbf{q},\omega)}. \qquad (15)$$

Inserting Eq. (14) in the macroscopic average of Eq. (10), we obtain for the macroscopic electric displacement vector

$$\mathbf{D}_0(\mathbf{q},\omega) = \mathbf{E}_0(\mathbf{q},\omega) + 4\pi A(\mathbf{q},\omega)\tilde{\alpha}_{0,0}^{(1)}(\mathbf{q},\mathbf{q},\omega)\mathbf{E}_0(\mathbf{q},\omega) + A(\mathbf{q},\omega)\mathbf{P}_0^{(2)}(\mathbf{q},\omega), \qquad (16)$$

which gives to first order the macroscopic dielectric function

$$\epsilon_M(\mathbf{q},\omega) = 1 + 4\pi\tilde{\alpha}_{0,0}^{(1)}(\mathbf{q},\mathbf{q},\omega)A(\mathbf{q},\omega) \qquad (17)$$

recovering the same results obtained in Ref.[51], as expected.

Besides this linear contribution for $\mathbf{D}$, we observe that Eq.(16) contains also the nonlinear term

$$\mathbf{D}_0^{(2)}(\mathbf{q},\omega) = 4\pi A(\mathbf{q},\omega)\mathbf{P}_0^{(2)}(\mathbf{q},\omega). \qquad (18)$$

We define the macroscopic second-order polarization $P_M^{(2)}$ as

$$\mathbf{D}_0^{(2)}(\mathbf{q},\omega) = 4\pi P_M^{(2)}(\mathbf{q},\omega), \qquad (19)$$

where $P_M^{(2)}$ is related to the second-order susceptibility $\chi^{(2)}$ through the relation in Eq. (1). From Eq. (18) we obtain an expression for $P_M^{(2)}$, calculating $\mathbf{P}_0^{(2)}$ from the spatial macroscopic average of Eq. (9) and writing its explicit dependence on the total electric field. We note that at this point we only need the linear relation (valid for any $\omega$) between the two fields $\mathbf{E}^P$ and $\mathbf{E}$

$$\mathbf{E}_0^P(\mathbf{q},\omega) = \left[1 + 4\pi\frac{\mathbf{q}\,\mathbf{q}}{q\,q}\frac{\tilde{\alpha}_{0,0}^{(1)}(\mathbf{q},\mathbf{q},\omega)}{1 - 4\pi\tilde{\alpha}_{0,0}^{(1),LL}(\mathbf{q},\mathbf{q},\omega)}\right]\mathbf{E}_0(\mathbf{q},\omega) \qquad (20)$$

because $\mathbf{E}^P$ appears already twice in Eq. (9).

We finally obtain the second-order macroscopic polarization

$$P_M^{(2)}(\mathbf{q},\omega) = \sum_{\mathbf{q}_1\mathbf{q}_2} \int d\omega_1 d\omega_2 \delta(\omega-\omega_1-\omega_2)\delta_{\mathbf{q},(\mathbf{q}_1+\mathbf{q}_2)} A(\mathbf{q},\omega)\tilde{\alpha}^{(2)}_{0,0,0}(\mathbf{q},\mathbf{q}_1,\mathbf{q}_2,\omega_1,\omega_2) \times$$
$$A(\mathbf{q}_1,\omega_1)A(\mathbf{q}_2,\omega_2)\mathbf{E_0}(\mathbf{q}_1,\omega_1)\mathbf{E_0}(\mathbf{q}_2,\omega_2), \quad (21)$$

and considering that the second-order susceptibility is defined by the relation

$$P_M^{(2)}(\mathbf{q},\omega) = \sum_{\mathbf{q}_1\mathbf{q}_2} \int d\omega_1 d\omega_2 \delta(\omega-\omega_1-\omega_2)\delta_{\mathbf{q},(\mathbf{q}_1+\mathbf{q}_2)} \chi^{(2)}(\mathbf{q},\mathbf{q}_1,\mathbf{q}_2,\omega,\omega_1,\omega_2) \times$$
$$\mathbf{E_0}(\mathbf{q}_1,\omega_1)\mathbf{E_0}(\mathbf{q}_2,\omega_2), \quad (22)$$

the macroscopic second-order susceptibility is

$$\chi^{(2)}(\mathbf{q},\mathbf{q}_1,\mathbf{q}_2,\omega,\omega_1,\omega_2) = \delta_{\mathbf{q},(\mathbf{q}_1+\mathbf{q}_2)} A(\mathbf{q},\omega)\tilde{\alpha}^{(2)}_{0,0,0}(\mathbf{q},\mathbf{q}_1,\mathbf{q}_2,\omega_1,\omega_2) A(\mathbf{q}_1,\omega_1)A(\mathbf{q}_2,\omega_2). \quad (23)$$

## III. THE OPTICAL LIMIT OF THE MACROSCOPIC SECOND-ORDER SUSCEPTIBILITY

We derived an expression for $\chi^{(2)}$ (Eq. (23)) which describes the interaction of a material with an electric field containing both longitudinal and transverse components. However, as we are interested in the low energy part of the SHG spectrum (photons typically below 15 eV) we will consider only the optical limit, i.e. $q \to 0$. In this case, we consider that the direction of $\mathbf{q}$ is no longer defined for an uniform field and the responses depend only on the polarization of the field[53]. Therefore, the calculation of the response functions, considering either a longitudinal or a transverse perturbation, leads to the same information. Hence, we will express the second-order susceptibility, Eq. (23), in terms of longitudinal quantities only which allows us to use the electron density as the fundamental physical quantity of Eq. (23). Consequently, the quantities $\tilde{\alpha}^{(1)}$ and $\tilde{\alpha}^{(2)}$ can be expressed only in terms of the density-density $\chi_{\rho\rho}$ and the density-density-density $\chi_{\rho\rho\rho}$ response functions. It is thus natural to develop our approach in TDDFT which has also the main advantage that many-body effects can be included straightforwardly[50,54] at lower computational cost with respect to other theoretical methods[24,55–58].

In this section, we consider the macroscopic second-order polarization given in Eq. (22) for SHG and we rewrite this expression in terms of longitudinal quantities only. In this way we find an expression for $\chi^{(2)}$ which can be computed in the TDDFT framework.

## A. THE OPTICAL LIMIT

Starting with the general expression of the macroscopic second-order polarization Eq. (22), we consider the case of a longitudinal electric field $\mathbf{E}$ which can be written in a cartesian frame as $\mathbf{E} = (E_x, E_y, E_z)$. Furthermore, as in the following we always consider the limit $q \to 0$, we define a unit vector $\hat{\mathbf{q}} = (\hat{q}_x, \hat{q}_y, \hat{q}_z)$ which has components along the electric field $\mathbf{E}$. This vector $\hat{\mathbf{q}}$ is used throughout this derivation when the limit $q \to 0$ is taken.

In particular, in this paper, we derive an expression for $\chi^{(2)}$ in TDDFT for systems with zinc-blende cubic symmetry, where only one independent tensor component $\chi^{(2)}_{xyz}$ is non zero[94]. In this case, since $\mathbf{q}_1 = \mathbf{q}_2$, we will indicate the momentum vector as $\mathbf{q}$.

We point out that this approach can be applied to any symmetry of the system and other symmetries will be presented in a forthcoming paper[59].

The macroscopic second-order polarization in the case of the cubic zinc-blende symmetry, can be written as

$$P^{(2)}_M(2\omega) = \begin{pmatrix} \chi^{(2)}_{xyz} E_y(\omega) E_z(\omega) \\ \chi^{(2)}_{yxz} E_x(\omega) E_z(\omega) \\ \chi^{(2)}_{zxy} E_x(\omega) E_y(\omega) \end{pmatrix} \quad (24)$$

with $\chi^{(2)}_{xyz} = \chi^{(2)}_{yxz} = \chi^{(2)}_{zxy}$.

For longitudinal fields $P^{(2)}_M$ depends only on the direction of the electric field $\mathbf{E}$ and the longitudinal second-order macroscopic polarization $\hat{\mathbf{q}} P^{(2)}_M(2\omega)$ is defined as

$$\hat{\mathbf{q}} P^{(2)}_M(2\omega) = 3\chi^{(2)}_{xyz} \hat{q}_x \hat{q}_y \hat{q}_z E^2(\omega). \quad (25)$$

In the following, in order to obtain an expression for $\chi^{(2)}_{xyz}$, we compare the macroscopic polarization given in Eq. (25) to the general definition Eq. (21) in the limit of longitudinal fields.

In Eq. (20) the quantity $\tilde{\alpha}^{(1)}_{0,0}$, which has the general form[95]

$$\tilde{\alpha}^{(1)}_{0,0}(\mathbf{q}, \mathbf{q}, \omega) = \begin{pmatrix} \tilde{\alpha}^{(1),LL}_{0,0}(\mathbf{q}, \mathbf{q}, \omega) & \tilde{\alpha}^{(1),LT}_{0,0}(\mathbf{q}, \mathbf{q}, \omega) \\ \tilde{\alpha}^{(1),TL}_{0,0}(\mathbf{q}, \mathbf{q}, \omega) & \tilde{\alpha}^{(1),TT}_{0,0}(\mathbf{q}, \mathbf{q}, \omega) \end{pmatrix} \quad (26)$$

can be expressed as

$$1 + 4\pi \frac{\mathbf{q}}{q} \frac{\mathbf{q}}{q} \frac{\tilde{\alpha}^{(1)}_{0,0}}{1 - 4\pi\tilde{\alpha}^{(1),LL}} = \begin{pmatrix} \epsilon^{LL}_M & 4\pi\epsilon^{LL}_M \tilde{\alpha}^{(1),LT}_{0,0} \\ 0 & 1 \end{pmatrix} \quad (27)$$

where
$$\epsilon_M^{LL}(\mathbf{q},\omega) = \frac{1}{1 - 4\pi\tilde{\alpha}_{0,0}^{(1),LL}(\mathbf{q},\mathbf{q},\omega)}. \tag{28}$$

like in Ref.[51].

Using Eq. (28) in Eq. (21) we obtain

$$\chi_{xyz}^{(2)} = \frac{1}{3\hat{q}_x\hat{q}_y\hat{q}_z}\epsilon_M^{LL}(\hat{\mathbf{q}},2\omega)\epsilon_M^{LL}(\hat{\mathbf{q}},\omega)\epsilon_M^{LL}(\hat{\mathbf{q}},\omega) \times$$
$$\hat{\mathbf{q}}\tilde{\alpha}_{0,0,0}^{(2)}(2\mathbf{q},\mathbf{q},\mathbf{q},\omega,\omega)\hat{\mathbf{q}}\hat{\mathbf{q}}, \tag{29}$$

where the tensor $\tilde{\alpha}^{(2)}$ still appears.

We calculate $\tilde{\alpha}^{(2)}$ considering that the microscopic longitudinal polarization

$$P^{(2)L}(2\mathbf{q},2\omega) = \hat{\mathbf{q}}\tilde{\alpha}_{0,0,0}^{(2)}(2\mathbf{q},\mathbf{q},\mathbf{q},\omega,\omega)\hat{\mathbf{q}}\hat{\mathbf{q}}(E^P)^2, \tag{30}$$

is related to the induced density through the continuity equation

$$P^{(2)L}(2\mathbf{q},2\omega) = \frac{i}{2q}\rho_{ind}^{(2)}(2\mathbf{q},2\omega) \tag{31}$$

and taking into account the relation between the induced density and the response function, we have

$$\hat{\mathbf{q}}\tilde{\alpha}_{0,0,0}^{(2)}(2\mathbf{q},\mathbf{q},\mathbf{q},\omega,\omega)\hat{\mathbf{q}}\hat{\mathbf{q}} = \frac{-i}{4}\chi_{\rho\rho\rho}(2\hat{\mathbf{q}},\hat{\mathbf{q}},\hat{\mathbf{q}},\omega,\omega). \tag{32}$$

Substituting Eq. (32) in Eq. (29), we finally obtain for $\chi_{xyz}^{(2)}$

$$\chi_{xyz}^{(2)} = \frac{-i}{24\hat{q}_x\hat{q}_y\hat{q}_z}\epsilon_M^{LL}(\hat{\mathbf{q}},2\omega)\epsilon_M^{LL}(\hat{\mathbf{q}},\omega)\epsilon_M^{LL}(\hat{\mathbf{q}},\omega) \times$$
$$\chi_{\rho\rho\rho}(2\hat{\mathbf{q}},\hat{\mathbf{q}},\hat{\mathbf{q}},\omega,\omega) \tag{33}$$

where

$$\chi_{\rho\rho\rho}(2\hat{\mathbf{q}},\hat{\mathbf{q}},\hat{\mathbf{q}},\omega,\omega) = \lim_{q\to 0}\frac{1}{q^3}\chi_{\rho\rho\rho}(2\mathbf{q},\mathbf{q},\mathbf{q},\omega,\omega). \tag{34}$$

In this form the $\chi_{xyz}^{(2)}$ depends on the density-density-density response function $\chi_{\rho\rho\rho}$ and on the macroscopic longitudinal dielectric functions $\epsilon_M^{LL}$, which have to be evaluated at the frequency of the incoming ($\omega$) and outgoing ($2\omega$) photons.

## B. Calculation of the $\chi_{\rho\rho\rho}$ in the TDDFT

The second-order response function $\chi_{\rho\rho\rho}$ is calculated in TDDFT through a second-order Dyson-like equation[47,50,60]

$$\left[1 - \chi_0^{(1)}(2\omega) f_{uxc}(2\omega)\right] \chi_{\rho\rho\rho}^{(2)}(2\omega, \omega) =$$
$$\chi_0^{(2)}(2\omega, \omega) \left[1 + f_{uxc}(\omega)\chi^{(1)}(\omega)\right] \left[1 + f_{uxc}(\omega)\chi^{(1)}(\omega)\right]$$
$$+ \chi_0^{(1)}(2\omega) g_{xc}(\omega)\chi^{(1)}(\omega)\chi^{(1)}(\omega), \tag{35}$$

where we explicitly omit the dependence on the **q** and **G**-vectors. In fact, Eq. (35) is rather complex and we believe that the compact form used here is easier for comprehension. We show the full dependence on **q** and **G**-vectors of Eq. (35) in Appendix B.

In Eq. (35) appears the quantity $f_{uxc}$ which is the sum of the bare-coulomb potential $u$ and of the exchange-correlation kernel $f_{xc} = \frac{\delta V_{xc}}{\delta \rho}$ and the quantity $g_{xc} = \frac{\delta^2 V_{xc}}{\delta \rho \delta \rho'}$. Moreover, $\chi^{(1)}(\omega)$ is the linear response function calculated via the Dyson-like equation

$$[1 - \chi_0^{(1)}(\omega) f_{uxc}(\omega)] \chi^{(1)}(\omega) = \chi_0^{(1)}(\omega), \tag{36}$$

and the functions $\chi_0^{(1)}(\omega)$ and $\chi_0^{(2)}(\omega)$ are the linear and second-order response functions in the IPA.

Among all the quantities appearing in Eq. (35) we would like to point out $\chi_0^{(2)}(\omega)$, because it plays a major role for the calculation of $\chi_{\rho\rho\rho}$. Its full dependence on **q** and **G**-vectors is $\chi_{0,\mathbf{G},\mathbf{G}_1,\mathbf{G}_2}^{(2)}(2\mathbf{q}, \mathbf{q}, \mathbf{q}, \omega)$ which makes this quantity more complex and more computationally demanding to calculate with respect to the linear $\chi_{0,\mathbf{G},\mathbf{G}_1}^{(1)}(\mathbf{q}, \mathbf{q}, \omega)$.

Eq. (35) gives a formally exact representation of the $\chi_{\rho\rho\rho}$ for an interacting system and, like in the linear TDDFT, the many-body interactions are rigorously and straightforwardly included via the $f_{xc}$ and $g_{xc}$ kernels. Depending on the kernels we use, we can obtain different levels of approximation in the description of the many-body interactions in $\chi_{\rho\rho\rho}$. Up to now, most of the *ab initio* calculations existing in literature were obtained within IPA, which we recover by setting $f_{xc} = 0$ and $g_{xc} = 0$ and considering only $\mathbf{G} = \mathbf{G}_1 = \mathbf{G}_2 = 0$. In this case the factors $1 + u\chi^{(1)}$ and $1 - \chi^{(0)}u$ of Eq. (35) compensate the $\epsilon_M^{LL}$ functions of Eq. (33), leading to the usual expression $\chi_{xyz}^{(2)} = \chi_0^{(2)}$.

Instead, if we consider also the **G**-vectors different from zero in Eq. (35), we can describe the crystal local-field effects of our systems in the random-phase approximation (RPA). Beyond

RPA we have to consider also the two kernels ($f_{xc}$ and $g_{xc}$), which describe the excitonic effects.

## C. Calculation of the $\chi^{(2)}_{0,\mathbf{G},\mathbf{G}_1,\mathbf{G}_2}(2\mathbf{q},\mathbf{q},\mathbf{q},\omega)$

The $\chi^{(2)}_0$ is the second-order response function in IPA, which in terms of the Bloch wavefunctions reads

$$\chi^{(2)}_{0,\mathbf{G},\mathbf{G}_1,\mathbf{G}_2}(2\mathbf{q},\mathbf{q},\mathbf{q},\omega) = \frac{2}{V} \sum_{n,n',n'',\mathbf{k}} \frac{\langle\phi_{n,\mathbf{k}}|e^{-i(2\mathbf{q}+\mathbf{G})\mathbf{r}}|\phi_{n',\mathbf{k}+2\mathbf{q}}\rangle}{(E_{n,\mathbf{k}} - E_{n',\mathbf{k}+2\mathbf{q}} + 2\omega + 2i\eta)}$$
$$\left[ (f_{n,\mathbf{k}} - f_{n'',\mathbf{k}+\mathbf{q}}) \frac{\langle\phi_{n',\mathbf{k}+2\mathbf{q}}|e^{i(\mathbf{q}+\mathbf{G}_1)\mathbf{r}_1}|\phi_{n'',\mathbf{k}+\mathbf{q}}\rangle \langle\phi_{n'',\mathbf{k}+\mathbf{q}}|e^{i(\mathbf{q}+\mathbf{G}_2)\mathbf{r}_2}|\phi_{n,\mathbf{k}}\rangle}{(E_{n,\mathbf{k}} - E_{n'',\mathbf{k}+\mathbf{q}} + \omega + i\eta)} \right.$$
$$+ (f_{n,\mathbf{k}} - f_{n'',\mathbf{k}+\mathbf{q}}) \frac{\langle\phi_{n',\mathbf{k}+2\mathbf{q}}|e^{i(\mathbf{q}+\mathbf{G}_2)\mathbf{r}_2}|\phi_{n'',\mathbf{k}+\mathbf{q}}\rangle \langle\phi_{n'',\mathbf{k}+\mathbf{q}}|e^{i(\mathbf{q}+\mathbf{G}_1)\mathbf{r}_1}|\phi_{n,\mathbf{k}}\rangle}{(E_{n,\mathbf{k}} - E_{n'',\mathbf{k}+\mathbf{q}} + \omega + i\eta)}$$
$$+ (f_{n',\mathbf{k}+2\mathbf{q}} - f_{n'',\mathbf{k}+\mathbf{q}}) \frac{\langle\phi_{n',\mathbf{k}+2\mathbf{q}}|e^{i(\mathbf{q}+\mathbf{G}_2)\mathbf{r}_2}|\phi_{n'',\mathbf{k}+\mathbf{q}}\rangle \langle\phi_{n'',\mathbf{k}+\mathbf{q}}|e^{i(\mathbf{q}+\mathbf{G}_1)\mathbf{r}_1}|\phi_{n,\mathbf{k}}\rangle}{(E_{n'',\mathbf{k}+\mathbf{q}} - E_{n',\mathbf{k}+2\mathbf{q}} + \omega + i\eta)}$$
$$\left. + (f_{n',\mathbf{k}+2\mathbf{q}} - f_{n'',\mathbf{k}+\mathbf{q}}) \frac{\langle\phi_{n',\mathbf{k}+2\mathbf{q}}|e^{i(\mathbf{q}+\mathbf{G}_1)\mathbf{r}_1}|\phi_{n'',\mathbf{k}+\mathbf{q}}\rangle \langle\phi_{n'',\mathbf{k}+\mathbf{q}}|e^{i(\mathbf{q}+\mathbf{G}_2)\mathbf{r}_2}|\phi_{n,\mathbf{k}}\rangle}{(E_{n'',\mathbf{k}+\mathbf{q}} - E_{n',\mathbf{k}+2\mathbf{q}} + \omega + i\eta)} \right], \quad (37)$$

where $f_{n,\mathbf{k}}$ are Fermi occupation numbers, the factor 2 accounts for the spin and $V$ is the volume of the cell. $\chi^{(2)}_{0,\mathbf{G},\mathbf{G}_1,\mathbf{G}_2}(2\mathbf{q},\mathbf{q},\mathbf{q},\omega)$ is a cube in the space of the $\mathbf{G}$-vectors, and in the limit $q \to 0$, we treat with special care the matrix elements for $\mathbf{G} = 0$ and/or $\mathbf{G}_1 = 0$ and/or $\mathbf{G}_2 = 0$. In the following we present only the $\mathbf{G} = \mathbf{G}_1 = \mathbf{G}_2 = 0$ component (head of the cube) of $\chi^{(2)}_0$, while for all the other cases we refer to Ref.[61]. We calculate Eq. (37) in the limit $q \to 0$ using $\mathbf{k} \cdot \mathbf{p}$ perturbation theory[62] in the case of degenerate states and when nonlocal Kleinman-Bylander pseudopotentials are used in the electronic structure calculation.

### 1. $\mathbf{k} \cdot \mathbf{p}$ *perturbation theory*

The Bloch wavefunctions $|\phi_{n,\mathbf{k}}\rangle$ and eigenvalues $E_{n,\mathbf{k}}$, appearing in Eq. (37), are solution of the Bloch hamiltonian $H_{\mathbf{k}}$[63]. In Eq. (37) appear also wavefunctions and eigenvalues corresponding to the wavevector $\mathbf{k} + \mathbf{q}$, solution of the Bloch hamiltonian $H_{\mathbf{k}+\mathbf{q}}$. Therefore, as we are considering the limit $q \to 0$, we use the $\mathbf{k} \cdot \mathbf{p}$ perturbation theory to expand Eq. (37). The first non-vanishing contribution in $\chi^{(2)}_0$ is of third order in terms of $q$ and we find that only the first-order energy correction and the first and the second-order wavefunction corrections contribute to the calculation of $\chi_0^{(2)}$.

We expand $H_{\mathbf{k}+\mathbf{q}}$ up to the second-order in the $\mathbf{q}$-vector, obtaining $H_{\mathbf{k}} + H_{\mathbf{k}}^{(1)} + H_{\mathbf{k}}^{(2)}$ where

$$H_{\mathbf{k}}^{(1)} = \mathbf{q}\mathbf{v} \quad ; \quad \mathbf{v} = \mathbf{p} + i[V_{nl}, \mathbf{r}] \tag{38}$$

and

$$H_{\mathbf{k}}^{(2)} = -\frac{i}{2}[\mathbf{q}\mathbf{r}, \mathbf{q}\mathbf{v}] = -\frac{i}{2}[\mathbf{q}\mathbf{r}, \mathbf{q}\mathbf{p} + i\mathbf{q}[V_{nl}, \mathbf{r}]], \tag{39}$$

where $V_{nl}$ is the nonlocal part of the ionic pseudopotential.

We obtain for the energy $E_{n,\mathbf{k}+\mathbf{q}} = E_{n,\mathbf{k}} + E_{n,\mathbf{k}}^{(1)}$ where

$$E_n^{(1)} = \langle \phi_{n,\mathbf{k}} | \mathbf{q}\mathbf{v} | \phi_{n,\mathbf{k}} \rangle, \tag{40}$$

and for the wavefunction we obtain $|\phi_{n,\mathbf{k}+\mathbf{q}}\rangle = e^{i\mathbf{q}\mathbf{r}}\left(|\phi_{n,\mathbf{k}}\rangle + |\phi_{n,\mathbf{k}}^{(1)}\rangle + |\phi_{n,\mathbf{k}}^{(2)}\rangle\right)$, where

$$|\phi_{n,\mathbf{k}}^{(1)}\rangle = \sum_{m \notin D_n} \frac{\langle \phi_{m,\mathbf{k}} | \mathbf{q}\mathbf{v} | \phi_{n,\mathbf{k}}\rangle}{E_{n,\mathbf{k}} - E_{m,\mathbf{k}}} |\phi_{m,\mathbf{k}}\rangle \tag{41}$$

and

$$|\phi_{n,\mathbf{k}}^{(2)}\rangle = \sum_{m,p \notin D_n} \left[ \frac{\langle \phi_{m,\mathbf{k}}|\mathbf{q}\mathbf{v}|\phi_{p,\mathbf{k}}\rangle \langle \phi_{p,\mathbf{k}}|\mathbf{q}\mathbf{v}|\phi_{n,\mathbf{k}}\rangle}{(E_{n,\mathbf{k}} - E_{p,\mathbf{k}})(E_{n,\mathbf{k}} - E_{m,\mathbf{k}})} - \right.$$
$$\langle \phi_{n,\mathbf{k}}|\mathbf{q}\mathbf{v}|\phi_{n,\mathbf{k}}\rangle \sum_{m \notin D_n} \frac{\langle \phi_{m,\mathbf{k}}|\mathbf{q}\mathbf{v}|\phi_{n,\mathbf{k}}\rangle}{(E_{n,\mathbf{k}} - E_{m,\mathbf{k}})^2} +$$
$$\left. \sum_{m \notin D_n} \frac{\langle \phi_{m,\mathbf{k}}| - \frac{i}{2}[\mathbf{q}\mathbf{r}, \mathbf{q}\mathbf{v}] |\phi_{n,\mathbf{k}}\rangle}{(E_{n,\mathbf{k}} - E_{m,\mathbf{k}})} \right] |\phi_{m,\mathbf{k}}\rangle -$$
$$\frac{1}{2}\sum_{m \notin D_n} \frac{|\langle \phi_{m,\mathbf{k}}|\mathbf{q}\mathbf{v}|\phi_{n,\mathbf{k}}\rangle|^2}{(E_{n,\mathbf{k}} - E_{m,\mathbf{k}})^2}|\phi_{n,\mathbf{k}}\rangle, \tag{42}$$

where $D_n$ corresponds to the degenerate subspace with $|\phi_{n,\mathbf{k}}\rangle$.

We recalculate Eq. (37) using the above expansion for the energy and wavefunction and in the case $\mathbf{G} = \mathbf{G}' = \mathbf{G}'' = \mathbf{0}$ we obtain that the second-order response function $\chi_0^{(2)}$ is the sum of two terms $\chi_{0,inter}^{(2)}$

$$\chi_{0,inter}^{(2)}(2\mathbf{q}, \mathbf{q}, \mathbf{q}, \omega, \omega) = \frac{4}{V} \sum_{n,n',n'',\mathbf{k}} \frac{\langle \phi_{n,\mathbf{k}}| - 2i\mathbf{q}\hat{\mathbf{r}}|\phi_{n',\mathbf{k}}\rangle}{(E_{n,\mathbf{k}} - E_{n',\mathbf{k}} + 2\omega + 2i\eta)} \times$$
$$\left[ (f_{n,\mathbf{k}} - f_{n'',\mathbf{k}}) \frac{\langle \phi_{n',\mathbf{k}}|i\mathbf{q}\hat{\mathbf{r}}|\phi_{n'',\mathbf{k}}\rangle \langle \phi_{n'',\mathbf{k}}|i\mathbf{q}\hat{\mathbf{r}}|\phi_{n,\mathbf{k}}\rangle}{(E_{n,\mathbf{k}} - E_{n'',\mathbf{k}} + \omega + i\eta)} + \right.$$
$$\left. (f_{n',\mathbf{k}} - f_{n'',\mathbf{k}}) \frac{\langle \phi_{n',\mathbf{k}}|i\mathbf{q}\hat{\mathbf{r}}|\phi_{n'',\mathbf{k}}\rangle \langle \phi_{n'',\mathbf{k}}|i\mathbf{q}\hat{\mathbf{r}}|\phi_{n,\mathbf{k}}\rangle}{(E_{n'',\mathbf{k}} - E_{n',\mathbf{k}} + \omega + i\eta)} \right] \tag{43}$$

and $\chi^{(2)}_{0,intra}$

$$\chi^{(2)}_{0,intra}(2\mathbf{q},\mathbf{q},\mathbf{q},\omega,\omega) = \frac{4}{V} \sum_{n,n',n'',\mathbf{k}} (f_{n,\mathbf{k}} - f_{n',\mathbf{k}}) \frac{\langle \phi_{n,\mathbf{k}} | -2i\mathbf{q}\hat{\mathbf{r}} | \phi_{n',\mathbf{k}} \rangle}{(E_{n',\mathbf{k}} - E_{n,\mathbf{k}})} \times$$

$$\frac{(2E_{n'',\mathbf{k}} - E_{n',\mathbf{k}} - E_{n,\mathbf{k}})}{E_{n',\mathbf{k}} - E_{n,\mathbf{k}}} \left[ 2\frac{\langle \phi_{n',\mathbf{k}} | i\mathbf{q}\hat{\mathbf{r}} | \phi_{n'',\mathbf{k}} \rangle \langle \phi_{n'',\mathbf{k}} | i\mathbf{q}\hat{\mathbf{r}} | \phi_{n,\mathbf{k}} \rangle}{(E_{n,\mathbf{k}} - E_{n',\mathbf{k}} + 2\omega + 2i\eta)} - \frac{1}{2} \frac{\langle \phi_{n',\mathbf{k}} | i\mathbf{q}\hat{\mathbf{r}} | \phi_{n'',\mathbf{k}} \rangle \langle \phi_{n'',\mathbf{k}} | i\mathbf{q}\hat{\mathbf{r}} | \phi_{n,\mathbf{k}} \rangle}{(E_{n,\mathbf{k}} - E_{n',\mathbf{k}} + \omega + i\eta)} \right] \quad (44)$$

where the operator $\hat{\mathbf{r}}$ is defined in terms of the position operator $\mathbf{r}$

$$\langle \phi_{n,\mathbf{k}} | \hat{\mathbf{r}} | \phi_{n',\mathbf{k}} \rangle = \begin{cases} \langle \phi_{n,\mathbf{k}} | \mathbf{r} | \phi_{n',\mathbf{k}} \rangle & \text{if } E_{n,\mathbf{k}} \neq E_{n',\mathbf{k}} \\ 0 & \text{if } E_{n,\mathbf{k}} = E_{n',\mathbf{k}}. \end{cases}$$

In practice the matrix element of $\mathbf{r}$ is calculated as the matrix element of the velocity operator $\mathbf{v}$

$$\langle \phi_{n,\mathbf{k}} | \mathbf{r} | \phi_{n',\mathbf{k}} \rangle = -i \frac{\langle \phi_{n,\mathbf{k}} | \mathbf{v} | \phi_{n',\mathbf{k}} \rangle}{E_{n,\mathbf{k}} - E_{n',\mathbf{k}}} \quad (45)$$

with $E_{n,\mathbf{k}} \neq E_{n',\mathbf{k}}$.

Finally, we point out that the fact to write $\chi^{(2)}_0 = \chi^{(2)}_{0,inter} + \chi^{(2)}_{0,intra}$ is completely arbitrary, but this separation is helpful in comparing with other theoretical formalism presented in literature[34,43].

2. *Inclusion of the scissors operator in the $\mathbf{k} \cdot \mathbf{p}$ perturbation theory*

To calculate the energies and wavefunctions appearing in the $\chi^{(2)}_{0,inter}$ (Eq. (43)) and $\chi^{(2)}_{0,intra}$ (Eq. (44)), we perform a Density-Functional Theory (DFT) in LDA calculation. However, in principle, electron bands have to be calculated within the many-body formalism using the $GW$ approach[24,64]. The application of this method to the calculation of the second-order response is not trivial and we prefer to use a simpler approach which can embody $GW$ gap corrections, the so-called scissors operator approximation[65].

In optical linear response, scissors operator approximation consists only in a rigid shift of the conduction bands, without changing the matrix elements and it has given excellent results for semiconductors[24]. In the second-order response the inclusion of the scissors operator is instead more complex than in the linear response as clearly pointed out by Nastos *et al*[66]. In our formalism, we explicitly include the scissors operator

$$S = \Delta \sum_{n\mathbf{k}} (1 - f_n) | \phi_{n,\mathbf{k}} \rangle \langle \phi_{n,\mathbf{k}} | \quad (46)$$

in the Bloch hamiltonian $H_{\mathbf{k}+\mathbf{q}}$ and we proceed as in Sec. III C 1.

We obtain for the first-order correction of the hamiltonian

$$H_{1,\mathbf{k}} = \mathbf{qv} + [S, i\mathbf{qr}], \tag{47}$$

and for the second-order correction

$$H_{2,\mathbf{k}} = -\frac{i}{2}[\mathbf{qr}, \mathbf{qv}] + \frac{1}{2}[\mathbf{qr}, [S, \mathbf{qr}]]. \tag{48}$$

For the expansion of the energy $E_{n,\mathbf{k}+\mathbf{q}}$ and the wavefunction $|\phi_{n,\mathbf{k}+\mathbf{q}}\rangle$, we find that the first-order correction of the energy does not change (see Eq.(40)) with respect to the calculation without the scissors operator, while the first-order correction of the wavefunction is

$$|\phi_{n,\mathbf{k}}^{(1)}\rangle = \sum_{m \notin D_n} \frac{\langle \phi_{m,\mathbf{k}}|\mathbf{qv}|\phi_{n,\mathbf{k}}\rangle}{E_{n,\mathbf{k}}^{SC} - E_{m,\mathbf{k}}^{SC}}|\phi_{m,\mathbf{k}}\rangle +$$
$$(f_{n,\mathbf{k}} - f_{m,\mathbf{k}})\frac{\langle \phi_{m,\mathbf{k}}|i\mathbf{qr}|\phi_{n,\mathbf{k}}\rangle}{E_{n,\mathbf{k}}^{SC} - E_{m,\mathbf{k}}^{SC}}|\phi_{m,\mathbf{k}}\rangle \tag{49}$$

and the second-order correction becomes

$$|\phi_{n,\mathbf{k}}^{(2)}\rangle = \sum_{m,p \notin D_n} \left[ \frac{\langle \phi_{m,\mathbf{k}}|\mathbf{qv}|\phi_{p,\mathbf{k}}\rangle\langle \phi_{p,\mathbf{k}}|\mathbf{qv}|\phi_{n,\mathbf{k}}\rangle}{(E_{n,\mathbf{k}}^{SC} - E_{p,\mathbf{k}}^{SC})(E_{n,\mathbf{k}}^{SC} - E_{m,\mathbf{k}}^{SC})} + \right.$$
$$(f_{p,\mathbf{k}} - f_{m,\mathbf{k}})(f_{n,\mathbf{k}} - f_{p,\mathbf{k}})\frac{\langle \phi_{m,\mathbf{k}}|i\mathbf{q\hat{r}}|\phi_{p,\mathbf{k}}\rangle\langle \phi_{p,\mathbf{k}}|i\mathbf{q\hat{r}}|\phi_{n,\mathbf{k}}\rangle}{(E_{n,\mathbf{k}}^{SC} - E_{p,\mathbf{k}}^{SC})(E_{n,\mathbf{k}}^{SC} - E_{m,\mathbf{k}}^{SC})} +$$
$$(f_{n,\mathbf{k}} - f_{p,\mathbf{k}})\frac{\langle \phi_{m,\mathbf{k}}|\mathbf{qv}|\phi_{p,\mathbf{k}}\rangle\langle \phi_{p,\mathbf{k}}|i\mathbf{q\hat{r}}|\phi_{n,\mathbf{k}}\rangle}{(E_{n,\mathbf{k}}^{SC} - E_{p,\mathbf{k}}^{SC})(E_{n,\mathbf{k}}^{SC} - E_{m,\mathbf{k}}^{SC})} +$$
$$\left. (f_{p,\mathbf{k}} - f_{m,\mathbf{k}})\frac{<\phi_{m,\mathbf{k}}|i\mathbf{q\hat{r}}|\phi_{p,\mathbf{k}}\rangle\langle \phi_{p,\mathbf{k}}|\mathbf{qv}|\phi_{n,\mathbf{k}}\rangle}{(E_{n,\mathbf{k}} - E_{p,\mathbf{k}})(E_{n,\mathbf{k}}^{SC} - E_{m,\mathbf{k}}^{SC})} \right] |\phi_{m,\mathbf{k}}\rangle -$$
$$\langle \phi_{n,\mathbf{k}}|\mathbf{qv}|\phi_{n,\mathbf{k}}\rangle \sum_{m \notin D_n} \left[ \frac{\langle \phi_{m,\mathbf{k}}|\mathbf{qv}|\phi_{n,\mathbf{k}}\rangle}{(E_{n,\mathbf{k}}^{SC} - E_{m,\mathbf{k}}^{SC})^2} + (f_{n,\mathbf{k}} - f_{p,\mathbf{k}})\frac{\langle \phi_{m,\mathbf{k}}|i\mathbf{q\hat{r}}|\phi_{n,\mathbf{k}}\rangle}{(E_{n,\mathbf{k}}^{SC} - E_{m,\mathbf{k}}^{SC})^2} + \right.$$
$$\left. \frac{\langle \phi_{m,\mathbf{k}}| -\frac{i}{2}[\mathbf{q\hat{r}}, \mathbf{qv}] + \frac{1}{2}[\mathbf{q\hat{r}}, [S, \mathbf{q\hat{r}}]]|\phi_{n,\mathbf{k}}\rangle}{(E_{n,\mathbf{k}}^{SC} - E_{m,\mathbf{k}}^{SC})} \right] |\phi_{m,\mathbf{k}}\rangle -$$
$$\frac{1}{2}\sum_{m \notin D_n}\left[\frac{|\langle \phi_{m,\mathbf{k}}|\mathbf{qv}|\phi_{n,\mathbf{k}}\rangle|}{(E_{n,\mathbf{k}}^{SC} - E_{m,\mathbf{k}}^{SC})^2}|\phi_{n,\mathbf{k}}\rangle + (f_{n,\mathbf{k}} - f_{m,\mathbf{k}})\frac{\langle \phi_{m,\mathbf{k}}|i\mathbf{q\hat{r}}|\phi_{n,\mathbf{k}}\rangle|^2}{(E_{n,\mathbf{k}}^{SC} - E_{m,\mathbf{k}}^{SC})^2}\right]|\phi_{n,\mathbf{k}}\rangle. \tag{50}$$

In this derivation the scissors operator does not act on the wavefunctions, thus letting them unchanged in the calculation.

Using these expansions to rewrite $\chi_0^{(2)}$ from Eq. (37), it turns out that the expression for $\chi^{(2)}_{0,inter}$ has the same form as Eq. (43), where the conduction energies have to be replaced by the shifted energies, while the valence energies are unchanged, exactly as in the linear response.

Instead, for the $\chi^{(2)}_{0,intra}$ we obtain

$$\chi^{(2)}_{0,intra}(2\mathbf{q},\mathbf{q},\mathbf{q},\omega,\omega) = \frac{4}{V} \sum_{n,n',n'',\mathbf{k}} (f_{n,\mathbf{k}} - f_{n',\mathbf{k}}) \frac{\langle \phi_{n,\mathbf{k}} | -2i\mathbf{q}\hat{\mathbf{r}} | \phi_{n',\mathbf{k}} \rangle}{(E^{SC}_{n',\mathbf{k}} - E^{SC}_{n,\mathbf{k}})} \times$$

$$\frac{(2E_{n'',\mathbf{k}} - E_{n',\mathbf{k}} - E_{n,\mathbf{k}})}{E_{n',\mathbf{k}} - E_{n,\mathbf{k}}} \left[ 2 \frac{\langle \phi_{n',\mathbf{k}} | i\mathbf{q}\hat{\mathbf{r}} | \phi_{n'',\mathbf{k}} \rangle \langle \phi_{n'',\mathbf{k}} | i\mathbf{q}\hat{\mathbf{r}} | \phi_{n,\mathbf{k}} \rangle}{(E^{SC}_{n,\mathbf{k}} - E^{SC}_{n',\mathbf{k}} + 2\omega + 2i\eta)} - \frac{1}{2} \frac{\langle \phi_{n',\mathbf{k}} | i\mathbf{q}\hat{\mathbf{r}} | \phi_{n'',\mathbf{k}} \rangle \langle \phi_{n'',\mathbf{k}} | i\mathbf{q}\hat{\mathbf{r}} | \phi_{n,\mathbf{k}} \rangle}{(E^{SC}_{n,\mathbf{k}} - E^{SC}_{n',\mathbf{k}} + \omega + 2i\eta)} \right] \quad (51)$$

where shifted ($E^{SC}$) and unshifted ($E$) energies are mixed.

We note that in this case, when the matrix element of $\hat{\mathbf{r}}$ is calculated as a matrix element of the operator $\mathbf{v}$ as in Eq. (45) the energies are not affected by the scissor correction.

## IV. RESULTS

In this section we present the SHG spectra for silicon carbide (SiC), aluminum arsenide (AlAs) and gallium arsenide (GaAs) semiconductors, which crystallize in zinc-blende structures. The calculations of the spectra are performed considering different levels of approximation of the many-body interactions, as explained in Sec. (III).

We first determine the electronic structure of the material in their ground-state with the Density-Functional Theory (DFT) in the LDA, using norm-conserving pseudopotentials[67] and plane-wave basis set with the ABINIT code[68]. For the calculation of the nonlinear optical spectra we use the nonlinear-response 2light code[69] implemented by us, on the basis of the linear-response Dp code[70].

Among the semiconductors studied, particular attention has been given to GaAs. In fact for Gallium the importance of the $d$ semicore states has been pointed out[71], therefore we perform our calculations using two different types of pseudopotentials for describing the electronic structure of this element. In particular, we use for all the calculations a pseudopotential which has the valence configuration $3d^{10}4s^24p^1$ and we compare, in the static limit, with a pseudopotential which does not contain the $d$ semicore states. In the following all presented results for GaAs are done using the pseudopotential with the $d$ semicore states, unless explicitly written.

The cutoff energies used are: 30 Ha for SiC and for AlAs and 50 Ha for GaAs, while for the GaAs calculations without the $d$ semicore states we use 20 Ha. All materials have been studied at their experimental lattice constant: 4.36 Å for SiC, 5.66 Å for AlAs and 5.65 Å for GaAs. Only for SiC we perform calculations using also the theoretical lattice constant of 4.33 Å. In order to simulate the quasiparticle energies, a scissors operators of 0.84 eV for SiC[72], 0.9 eV for AlAs[73] and 0.8 eV for GaAs[73] have been used. The spectra for SiC, AlAs and GaAs have been obtained using respectively 4096, 4096 and 17576 off-symmetry shifted $k$ points in the Brillouin zone (BZ) and the number of unoccupied states included in the calculation of the response functions is 6, 12 and 7. Crystal local-field effects are fully taken into account by carefully converging the size of all matrices in $(\mathbf{G}, \mathbf{G}_1, \mathbf{G}_2)$ space using 89 for SiC, 59 AlAs and 65 for GaAs number of **G**-vectors.

### A. The Independent Particle Approximation in the SHG spectra

The lowest level of approximation in the description of the electron-electron interaction is the IPA. In literature, even in this simple approximation, many discrepancies exist between different theoretical calculations and in some case the comparison with the experiments is only qualitative. The main difficulty in the calculation of the $\chi^{(2)}$ is its great sensitivity to the band structure of the material studied. In fact, many variables used for the theoretical description of the system, like for example lattice constants, pseudopotentials and values of the scissors correction, can influence the calculated band structure. A very small difference in the band structures used for constructing $\chi^{(2)}$ can induce large variations in its values, being of the order of 50% for some materials[30].

In Figs. (1,2,3) we show $\chi^{(2)}_{xyz}$ for SiC, AlAs and GaAs with (dashed) and without (solid) the scissors correction (see Eq (44), Eq (43) and Eq (51)). A scissors correction causes a blueshift in the spectrum, as well as a redistribution of the spectral weights, resulting in an overall decreasing of the intensity, in agreement with other calculations[43,44,46,66]. In Tab. (I) the static values of the $\chi^{(2)}_{xyz}$ are given for three values of the scissors correction. A small change of 0.1 eV in the value of the scissors correction can induce change of the order 4%-8% in the static second-order susceptibility, depending on the materials. In particular for Gallium, the design of the pseudopotential is not trivial and the importance of $d$ semicore states is well known[71]. The presence or the absence of the $d$ semicore states of Gallium influences the static value of $\chi^{(2)}_{xyz}$ for GaAs, as shown in Tab. (I). Using a pseudopotential with the $d$ semicore states we obtain higher values comparing with a pseudopotential without these states. The same effect was observed by Dal Corso et al.[41] including the non-linear core corrections in the pseudopotential and by Cabellos et al.[38] who performed an all-electron calculation.

Finally, we want to point out that also a small variation on the lattice constant can strongly influence the results. Levine et al.[30] showed for GaAs that using LDA determined lattice constant (5.493 Å) gives a $\chi^{(2)}_{xyz}$ of 160 pm/V which is 50% smaller than the one calculated using the experimental lattice constant (5.65 Å) of 358 pm/V. For the same material Dal Corso et al.[41] obtained a smaller discrepancy (20%) between the $\chi^{(2)}_{xyz}$ calculated with the two lattice constants. However, in this case their theoretical lattice constant (5.556 Å) is closer to the experimental 5.65Å. The big difference observed by Levine et al. for GaAs, is in fact due to the larger difference between the values of the lattice constants used. For AlAs, they found 68 pm/V for $\chi^{(2)}_{xyz}$ using the theoretical lattice constant (5.604 Å) and 78 pm/V using the experimental 5.66 Å. We confirm the same trend for SiC, for which we obtain 19.02 pm/V using the theoretical lattice constant (4.33 Å) and 20.04 pm/V using the experimental 4.36 Å.

### B. Crystal local-field effects on the SHG spectra

To go beyond the IPA, we have included the crystal local-field effects (RPA) in the calculation of the second-order susceptibility $\chi^{(2)}$, as in Eq. (35), setting the kernels $f_{xc} = 0$ and $g_{xc} = 0$. In this study, the spectra of SiC, AlAs and GaAs have been computed without scissors correction and they are shown in Figs. (4,5,6). The overall effect of the crystal local fields is to decrease the intensity of the second-order susceptibility with respect to IPA, without changing the shape of the spectra. Depending on the material and on the frequency range considered, the $\chi^{(2)}$ can decrease up to 30%. In the static limit this effect is of the order of the 15% for all the three semiconductors, as shown in Tabs. (II,III,IV), in agreement with other theoretical calculations[33,45].

## C. Excitonic effects on the SHG spectra

The exchange-correlation kernels $f_{xc}$ and $g_{xc}$ appearing in Eq. (35) take into account all dynamical exchange and correlation effects in the response of a system to an external perturbation, like excitonic effects. The main problem still remains to find a good approximation for the kernels. Furthermore, even if in the last ten years a big effort has been made towards the design of new efficient and sophisticated kernels, they have been applied only to linear optics. Therefore, it is not obvious whether these kernels can correctly describe the electron-hole correlation and exchange in the second-order process.

We have first considered the most widely used approximation for $f_{xc}$, the adiabatic local density approximation (ALDA)

$$f_{xc}^{ALDA}(\mathbf{r},\mathbf{r}',t,t') = \delta(t-t')\delta(\mathbf{r}-\mathbf{r'}) \left. \frac{\delta v_{xc}^{LDA}[\rho(\mathbf{r},\omega)]}{\delta \rho(\mathbf{r'},\omega)} \right|_{\omega=0}, \tag{52}$$

keeping $g_{xc}=0$.

Concerning the linear response, TDDFT with the ALDA kernel (TDLDA) has been demonstrated to yield good result in the linear response, especially for finite systems and electron energy loss spectra (EELS) of solids[74,75]. However, it is not sufficient to yield good absorption spectra in solids[24].

In Figs. (4,5,6) IPA, RPA and TDLDA are compared for SiC, AlAs and GaAs. When using the ALDA kernel the result remains very close to those of IPA and RPA. This is very similar to the behavior of TDLDA for the absorption spectra in solids[76], related to the lack of long-range contribution in the ALDA kernel[24]. To solve this issue, a model static long-range kernel has been proposed[73,77]

$$f_{xc}^{LRC} = -\frac{\alpha}{4\pi|\mathbf{r}-\mathbf{r'}|} \tag{53}$$

where $\alpha$ is a mean value for the dynamical dependence of $f_{xc}$, in a given range of frequency. This static long-range kernel yields a good total $f_{xc}$ if used on top of a GW or scissors corrected band structure and it has been shown to simulate correctly strong continuum excitons, i.e. materials with moderate electron-hole interactions, among which SiC, AlAs and GaAs. For more complex systems or stronger excitons the long-range kernel may fail. However this does not pose any limitation to our method as many-body effects can be included in TDDFT via more efficient and sophisticated kernels.

Here, we discuss the effects of the static long-range contribution $\alpha/q^2$ to the exchange-correlation kernel $f_{xc}$ for the second-order susceptibility, shown in Figs. (7,8,9). The values of $\alpha$ used are 0.2 for GaAs, 0.35 for AlAs and 0.5 for SiC as reported in Ref.[73]. The main effect of the $\alpha/q^2$ kernel is to increase the magnitude of $|\chi^{(2)}_{xyz}|$. This behavior can be understood by solving analytically Eq. (35) without crystal local-field effects, showing that the increase from the $|\chi^{(2)}_0|$ to the $|\chi^{(2)}_{xyz}|$ is proportional to $[1+\alpha/4\pi(\epsilon^{LL}_M(\omega)-1)]^2[1+\alpha/4\pi(\epsilon^{LL}_M(2\omega)-1)]$[47]. In Fig. (9) we also report the experimental second-harmonic generation spectrum measured by Bergfeld and Daum[78] (circles) for GaAs cubic semiconductor. In a previous work[47], we have pointed out that already within IPA we obtain the same shape as the experimental spectrum, in agreement with other theoretical studies Ref.[37,38]. However, within IPA the agreement with experiment is only qualitative and even though the shape of the theoretical spectrum is good, the relative intensity of the peaks and in particular the magnitude of the susceptibility is not in agreement with the experimental values. Moreover, as the inclusion of the crystal local-field effects cause a decreasing of the overall intensity of the spectrum, they do not improve the comparison with the experiment. Clearly, the IPA and the RPA are not sufficient to describe the physics of the process, and neither the TDLDA.

We have found[47] that only when excitonic effects are considered using the long-range kernel, an improvement in the description of the experimental data is achieved. In particular, we have obtained remarkable agreement in the experimental intensity of the highest peak of the second-order susceptibility. This result shows that a good description of the excitons is essential to correctly describe this nonlinear process, where the role of the kernel is crucial. Howewer, in Fig. (9) there are still some discrepancies between the experimental (circles) and our theoretical spectrum (solid), in particular in the low frequency range. This behavior shows the limit of the non-frequency dependent "long-range" kernel.

In order to investigate which is the origin of this difference, we have[47] more accurately described the macroscopic frequency-dependent dielectric function $\epsilon^{LL}_M$ appearing in Eq. (33). We have taken the experimental[79] dielectric function and we have used it in Eq. (33). In this way, we have achieved a better agreement, Fig. (10) (solid line), showing the importance of very accurate description of the exchange-correlation interactions.

The inclusion of non-locality, frequency dependence, memory effects in the kernel can improve the theoretical description of the $|\chi^{(2)}_{xyz}|$. Many efforts have been made in this direction for the linear response, for example, from the time-dependent current-density functional theory[80], from the Bethe Salpeter approach[24,81,82] and exact-exchange kernel approaches[83]. However, some of these kernels are too cumbersome to be applied to our approach and more specific studies are needed for the second-order response. In Fig. (10) the calculated spectrum (dashed) of Leitsman et al.[46] is also reported. They use excitonic states, obtained by the diagonalization of the effective Hamiltonian of the BSE approach, to construct the second-order susceptibility $|\chi^{(2)}_{xyz}|$. Even if they obtain a qualitative comparison with experimental data. some important discrepancies still remains.

The difference between our and Leitsman et al.[46] approach is mainly due to the fact that we use an exact Dyson equation, where we approximate the kernels, while Leitsman et al.[46] approximate particle-hole wavefunctions, leading to different results. Furthermore, another important point is how crystal local-field effects are treated in the two theoretical methods. In fact, Leitsman et al.[46] consider the crystal local fields only in the Coulomb exchange term of the effective Hamiltonian, while Eq. (33) shows that the connection between microscopic and macroscopic quantities is more complex.

### D. The static limit

In Tabs. (II,III,IV) we compare for SiC, AlAs and GaAs, the $|\chi^{(2)}_{xyz}|$ in the zero frequency limit with other theoretical calculations and where possible with experimental data. Among these results, quite large discrepancies exist, which are mainly due to the great sensitivity of this quantity to the calculation of the band structure, as explained in Sec. (IV A). The discrepancy in the theoretical calculations can be assigned to the different parameters used. However, all the theoretical calculations have the same relative trends considering the scissors correction, the crystal local-field and the excitonic effects. Both the scissors correction and the crystal local-field effects decreases the value of the $|\chi^{(2)}_{xyz}|$. In particular, the scissors correction decrease the $|\chi^{(2)}_{xyz}|$ of the order of 50% for GaAs, 40% for AlAs and 25% for SiC, while the effect of the crystal local-field is of the order of 15% for all the three semiconductors. When comparing our theoretical results with the experimental ones, we observe a large degree of discrepancy also for the experimental values. In fact, for GaAs, at photon energy of 0.117 eV Levine and Bethea[84] obtained 180±10 pm/V, later Roberts[85] recommended the value of 166 pm/V and more recently Eyres et al.[86] measured 172 pm/V at 0.118 eV. For AlAs we find the experimental value of 78±20[87].

## V. CONCLUSIONS

In conclusion, in this work we present the detailed derivation of the ab initio formalism[47] we develop for the calculation of the frequency-dependent second-order susceptibility $\chi^{(2)}$. This approach is valid for any kind of crystals and molecular systems and is based on TDDFT approach. In this way the crystal local-field and the excitonic effects are straightforwardly included in $\chi^{(2)}$. In our formalism, we first derive a relation between the microscopic and macroscopic formulation of the second-order responses. This allows us to obtain a general expression for the macroscopic second-order susceptibility $\chi^2$ valid for any fields (longitudinal and transverse). Then, we consider only the case of vanishing light wave vector ($q \to 0$), as we apply this formalism to the SHG spectroscopy. In this long wavelength limit, we rewrite the general expression for $\chi^2$ only for longitudinal fields, which permits us to write our formalism in the TDDFT framework. In our approach, in order to obtain a specific component of the tensor $\chi^2$, we explicitly consider the symmetry properties of the system. As we apply our method to the calculation of SHG spectroscopy for the cubic semiconductors SiC, AlAs and GaAs, we show in this paper the equation for the component $\chi^2_{xyz}$ for cubic symmetry. However, we point out that our approach can be applied to any symmetry of the system studied. We discuss SHG spectra obtained within different levels of description of the many-body interactions, starting from the Independent-Particle Approximation, we include quasiparticle effects via the scissors operator, crystal local-field and excitonic effects. In particular, we consider two different types of kernels: the ALDA and the "long-range" kernel. We find good agreement with other theoretical calculations and experiments presented in literature, showing the importance of very accurate description of the many-body interactions.

## VI. ACKNOWLEDGEMENTS

We wish to thank L. Reining, Ch. Giorgetti, F. Bechstedt, A. Rubio and F. Sottile for helpful discussions. This work was supported by the European Union through the

## APPENDIX A: DEFINITION OF THE RESPONSE FUNCTIONS

In section II we define the linear and second-order quasipolarizability: $\tilde{\alpha}^{(1)}$ and $\tilde{\alpha}^{(2)}$ defined in terms of the current/density response functions. The definitions of these response functions are[52]

$$\chi_{\mathbf{jj}}(\mathbf{r},\mathbf{r}',t-t') = i\theta(t-t')\langle\left[\hat{\mathbf{j}}_I(\mathbf{r},t),\hat{\mathbf{j}}_I(\mathbf{r}',t')\right]\rangle$$

$$\chi_{\mathbf{j}\rho}(\mathbf{r},\mathbf{r}',t-t') = i\theta(t-t')\langle\left[\hat{\mathbf{j}}_I(\mathbf{r},t),\hat{\rho}_I(\mathbf{r}',t')\right]\rangle \tag{A1}$$

$$\chi_{\rho\mathbf{j}}(\mathbf{r},\mathbf{r}',t-t') = i\theta(t-t')\langle\left[\hat{\rho}_I(\mathbf{r},t),\hat{\mathbf{j}}_I(\mathbf{r}',t')\right]\rangle \tag{A2}$$

$$\chi_{\rho\rho}(\mathbf{r},\mathbf{r}',t-t') = i\theta(t-t')\langle[\hat{\rho}_I(\mathbf{r},t),\hat{\rho}_I(\mathbf{r}',t')]\rangle \tag{A3}$$

$$\chi_{\mathbf{jjj}}(\mathbf{r},\mathbf{r}',\mathbf{r}'',t-t',t-t'') = \theta(t-t')\theta(t-t'')$$
$$T\langle\left[\left[\hat{\mathbf{j}}_I(\mathbf{r},t),\hat{\mathbf{j}}_I(\mathbf{r}',t')\right],\hat{\mathbf{j}}_I(\mathbf{r}'',t'')\right]\rangle. \tag{A4}$$

## APPENDIX B: THE SECOND-ORDER DYSON EQUATION FOR THE $\chi_{\rho\rho\rho}$

The longitudinal second-order response function $\chi_{\rho\rho\rho}$ can be calculated in TDDFT through the following second-order Dyson equation[47,50,60] written here in the reciprocal space

$$\sum_{G_2}\left[\delta_{G,G_2} - \sum_{G_1}\chi^{(1)}_{0,\mathbf{GG}_1}(2\mathbf{q},2\mathbf{q},2\omega)f_{uxc,\mathbf{G}_1\mathbf{G}_2}(2\mathbf{q},2\mathbf{q},2\omega)\right]\times$$

$$\chi_{\rho\rho\rho,\mathbf{G}_2\mathbf{G'}\mathbf{G''}}(2\mathbf{q},\mathbf{q},\mathbf{q},\omega) = \sum_{G_1 G_3}\chi^{(2)}_{0,\mathbf{GG}_1\mathbf{G}_3}(2\mathbf{q},\mathbf{q},\mathbf{q},2\omega)\times$$

$$\left[\delta_{\mathbf{G}_3\mathbf{G''}} + \sum_{G_4}f_{uxc,\mathbf{G}_3\mathbf{G}_4}(\mathbf{q},\mathbf{q},\omega)\chi^{(1)}_{\mathbf{G}_4\mathbf{G''}}(\mathbf{q},\mathbf{q},\omega)\right]\times$$

$$\left[\delta_{\mathbf{G'G}_1} + \sum_{\mathbf{G}_2}f_{uxc,\mathbf{G}_1\mathbf{G}_2}(\mathbf{q},\mathbf{q},\omega)\chi^{(1)}_{\mathbf{G}_2\mathbf{G'}}(\mathbf{q},\mathbf{q},\omega)\right]\times$$

$$\sum_{G_1,G_2 G_3}\chi^{(1)}_{0,\mathbf{GG}_1}(2\mathbf{q},2\mathbf{q},2\omega)g_{xc,\mathbf{G}_1\mathbf{G}_2\mathbf{G}_3}(2\mathbf{q},\mathbf{q},\mathbf{q},\omega)\times$$

$$\chi^{(1)}_{\mathbf{G}_2\mathbf{G'}}(\mathbf{q},\mathbf{q},\omega)\chi^{(1)}_{\mathbf{G}_3\mathbf{G''}}(\mathbf{q},\mathbf{q},\omega), \tag{B1}$$

where $\mathbf{q}$ is a vector in the first Brillouin zone, $\mathbf{G}$, $\mathbf{G}_1$ and $\mathbf{G}_2$ are vectors of the reciprocal lattice.

---

and $\tilde{\alpha}_{\mathbf{0},\mathbf{0}}^{(1),TL}$ are 2-dimension vectors and $\tilde{\alpha}_{\mathbf{0},\mathbf{0}}^{(1),TT}$ is a $2 \times 2$ matrix.

**FIGURES**

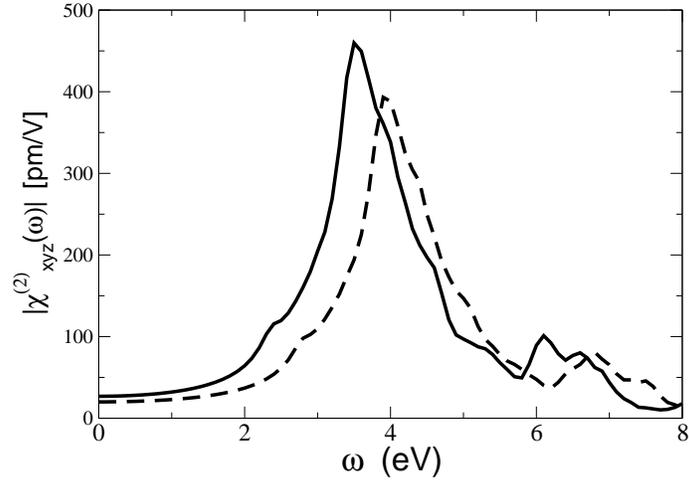

FIG. 1: $|\chi_{xyz}^{(2)}|$ for SiC. Solid line: IPA calculation. Dashed line: IPA plus scissors correction calculation, $\Delta = 0.84$ eV.

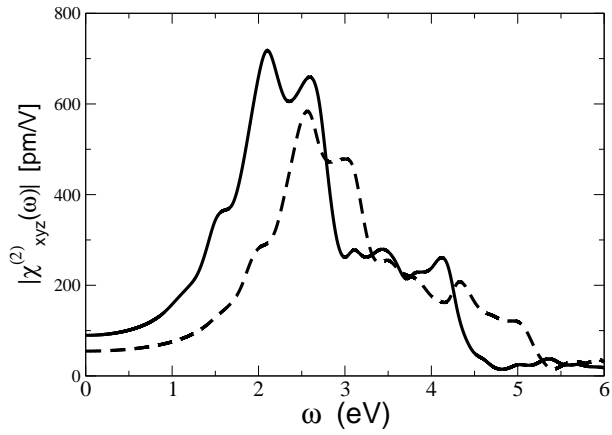

FIG. 2: $|\chi_{xyz}^{(2)}|$ for AlAs. Solid line: IPA calculation. Dashed line: IPA plus scissors correction calculation, $\Delta = 0.9$ eV.

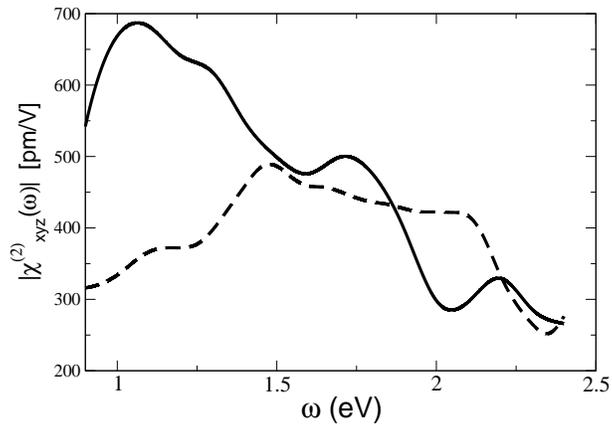

FIG. 3: $|\chi^{(2)}_{xyz}|$ for GaAs. Solid line: IPA calculation. Dashed line: IPA plus scissors correction calculation, $\Delta = 0.8$ eV.

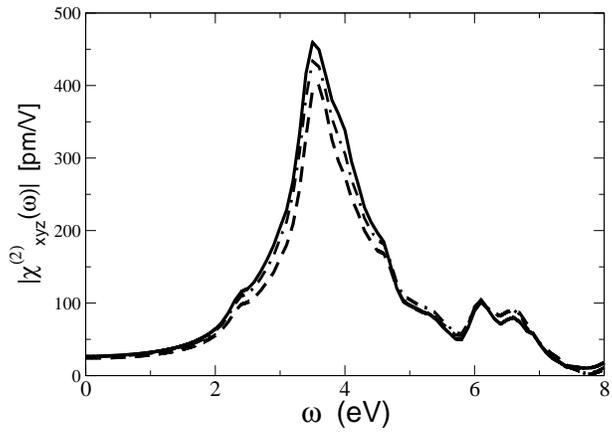

FIG. 4: $|\chi^{(2)}_{xyz}|$ for SiC. Solid line: IPA calculation. Dashed line: RPA calculation. Dot-dashed line: TDLDA calculation.

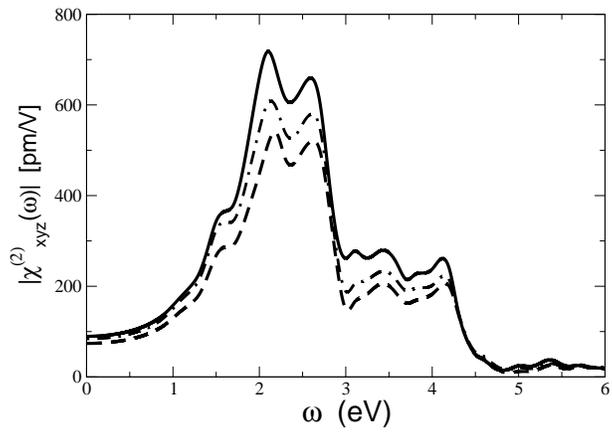

FIG. 5: $|\chi^{(2)}_{xyz}|$ for AlAs. Solid line: IPA calculation. Dashed line: RPA calculation. Dot-dashed line: TDLDA calculation.

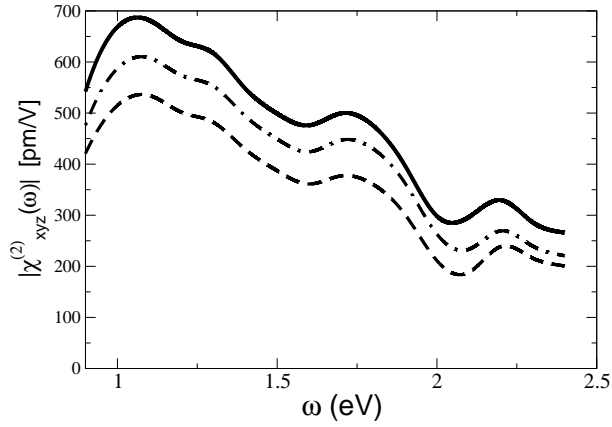

FIG. 6: $|\chi^{(2)}_{xyz}|$ for GaAs. Solid line: IPA calculation. Dashed line: RPA calculation. Dot-dashed line: TDLDA calculation.

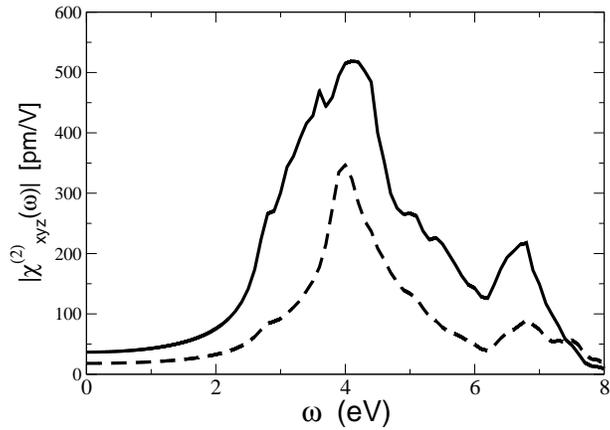

FIG. 7: $|\chi^{(2)}_{xyz}|$ for SiC. Dashed line: RPA calculation. Solid line: excitonic calculation using the $\alpha$ kernel plus scissors correction. For this material we have used $\alpha = 0.5$ and $\Delta = 0.84$ eV.

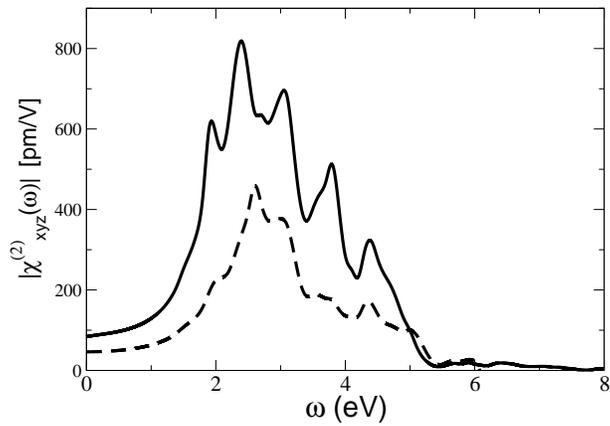

FIG. 8: $|\chi^{(2)}_{xyz}|$ for AlAs. Dashed line: RPA calculation. Solid line: excitonic calculation using the $\alpha$ kernel plus scissors correction. For this material we have used $\alpha = 0.35$ and $\Delta = 0.9$ eV.

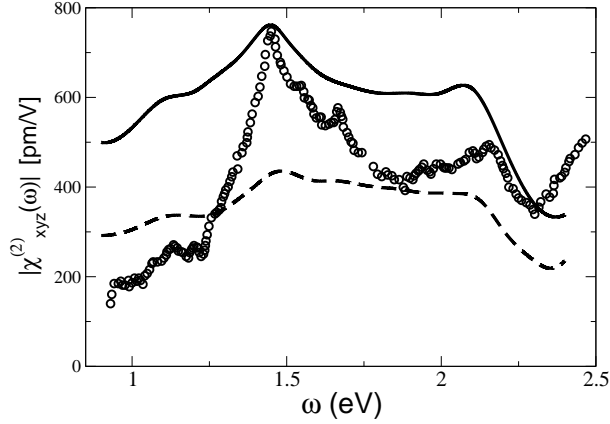

FIG. 9: $|\chi^{(2)}_{xyz}|$ for GaAs. Dashed line: RPA calculation. Solid line: excitonic calculation using the $\alpha$ kernel plus scissors correction. For this material we have used $\alpha = 0.2$ and $\Delta = 0.8$ eV. Circles: experimental spectrum from Bergfeld and Daum[78].

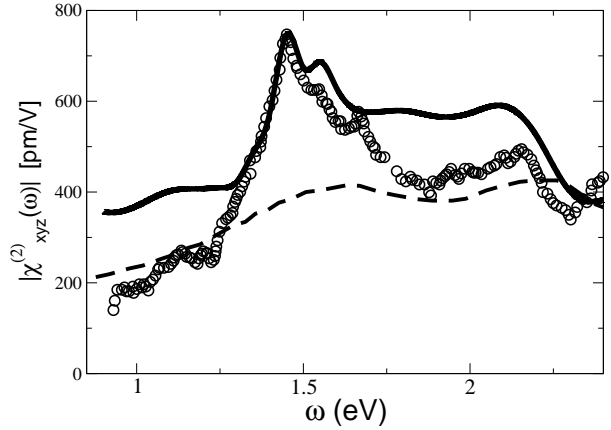

FIG. 10: $|\chi^{(2)}_{xyz}|$ for GaAs. Solid line: excitonic calculation using the $\alpha$ kernel plus scissors correction. For this material we have used $\alpha = 0.2$ and $\Delta = 0.8$ eV. In this case, to calculate $|\chi^{(2)}_{xyz}|$ as in Eq. (33) we have used the experimental $\epsilon_M^{LL}$[79]. Dashed line: Leitsmann *et al.*[46] where the excitons are included within BSE framework. Circles: experimental spectrum from Bergfeld and Daum[78].

**TABLES**

|       | scissors correction |        |        |
|-------|--------|--------|--------|
|       | 0.7 eV | 0.8 eV | 0.9 eV |
| SiC   | 20.99  | 20.13  | 19.64  |
| GaAs  | 244.81 | 224.39 | 206.65 |
| GaAs* | 224.19 | 208.03 | 193.59 |

TABLE I: $|\chi^{(2)}_{xyz}|$ [pm/V] for three different values of the scissors correction for SiC and GaAs. For Gallium we use two different pseudopotentials with and without the $d$ semicore states (with an asterisk).

| | SiC | | | | | |
|---|---|---|---|---|---|---|
| | IPA | IPA+SO | RPA | RPA+SO | ALDA | $\alpha$+SO |
| This work | 26.91 | 20.04 | 23.96 | 18.07 | 25.98 | 36.71 |
| Adolph et al.[43] | 26 | | 23 | | | |
| Chen et al.[33] | 24.8 | | 16 | | | |
| Rashkeev al.[44] | 17.6 | | 11.3 | | | |

TABLE II: Comparison of $|\chi^{(2)}_{xyz}|$ [pm/V] in the static limit with other theoretical calculations for SiC.

| | AlAs | | | | | |
|---|---|---|---|---|---|---|
| | IPA | IPA+SO | RPA | RPA+SO | ALDA | $\alpha$+SO |
| This work | 89.68 | 54.80 | 73.52 | 46.00 | 84.04 | 100.50 |
| Levine and Allan[30,31] | | 51 | 78 | 47.2 | | |
| Chang et al.[45] | | 55.8 | | | | 68.2 |
| Dal Corso et al.[41] | 64 | | | | | |
| Huang-Ching[88] | | 46 | | | | |
| Chen et al.[33] | 70.4 | 43.4 | 63.36 | 39.06 | | |
| Veithen et al.[89] | 70 | 42 | | | | |
| Souza et al.[90] | 64 | | | | | |
| Roman et al.[91] | 79 | | | | | |

TABLE III: Comparison of $|\chi^{(2)}_{xyz}|$ [pm/V] in the static limit with other theoretical calculations for AlAs.

|  | GaAs | | | | | |
|---|---|---|---|---|---|---|
|  | IPA | IPA+SO | RPA | RPA+SO | ALDA | $\alpha$+SO |
| This work | 559.28 | 224.39 | 480.85 | 192.04 | 522.70 | 216.54 |
| This work (no d) | 427.98 | 208.03 | 387.89 | 189.26 | 445.52 | 213.49 |
| Levine et al.[30,31] |  | 186.8 | 348 | 172.8 |  |  |
| Nastos et al.[66] |  | 206.6 |  |  |  |  |
| Chang et al.[45] |  | 196.4 |  |  |  | 236.4 |
| Rashkeev al.[44] | 735.6 | 162.0 |  |  |  |  |
| Dal Corso et al.[41] |  |  | 205 |  |  |  |
| Levine et al.[29] |  | 184.6 | 354 | 172.4 |  |  |
| Chen et al.[33] | 410.0 | 189.2 | 369.0 | 170.3 |  |  |
| Cabellos et al.[38] |  | (172.4)135.6 |  |  |  |  |
| Huang-Ching[88] |  | 251 |  |  |  |  |

TABLE IV: Comparison of $|\chi^{(2)}_{xyz}|$ [pm/V] in the static limit with other theoretical calculations for GaAs. We also show the results obtained not expliciting including the 3d semicore states of Gallium in the valence. For Cabellos *et al.*[38] we report in parenthesis the result obtained through an all-electron calculation.